\documentclass[prc,amsmath,showpacs,twocolumn]{revtex4}

\usepackage{graphicx}
\usepackage[dvips]{color}

\begin{document}

\title{Effective operators within the \textit{ab initio} no-core shell model}
\author{Ionel Stetcu}
\author{Bruce R. Barrett}
\affiliation{Department of Physics, University of Arizona, P.O. Box 210081, Tucson, Arizona 85721}
\author{Petr Navr\'atil}
\affiliation{Lawrence Livermore National Laboratory, Livermore, P.O. Box 808, California 94551}
\author{James P. Vary}
\affiliation{Department of Physics and Astronomy, Iowa State University, Ames, Iowa 50011}

\date{\today}

\begin{abstract}
We implement an effective operator formalism for general one- and two-body operators, obtaining results consistent with the no-core shell model (NCSM) wave functions. The Argonne V8' nucleon-nucleon potential was used in order to obtain realistic wave functions for $^4$He, $^6$Li and $^{12}$C. In the NCSM formalism, we compute electromagnetic properties using the two-body cluster approximation for the effective operators and obtain results which are sensitive to the range of the bare operator. To illuminate the dependence on the range, we employ a Gaussian two-body operator of variable range, finding weak renormalization of long range operators (e.g., quadrupole) in a fixed model space. This is understood in terms of the two-body cluster approximation which accounts mainly for short-range correlations. Consequently, short range operators, such as the relative kinetic energy, will be well renormalized in the two-body cluster approximation.
\end{abstract}
\pacs{21.60.Cs, 23.20.-g, 23.20.Js}
\maketitle

\section{Introduction}

The no-core shell model (NCSM) is a successful \textit{ab initio} method used to compute properties of light nuclei, starting from realistic two- and three- nucleon interactions. In this approach, all nucleons are active, but the Hilbert space available is a finite model space. In order to take into account the effect of the configurations outside the model space, we employ effective interactions obtained by means of an unitary transformation \cite{okubo,LS80,UMOA}, aimed to reproduce the low-lying spectra in our finite model space. Large basis NCSM calculations have been successful in the description of energy spectra in light nuclei \cite{c12lett,c12,spectr6}. However, the $E2$ transition strengths, powerful tests of the theoretical wave functions, are usually underestimated \cite{c12,spectr6}. To date, they were obtained with the bare operators, and one expects that a renormalization in the same fashion as for the interaction will give a better description. The only such effective operators computed so far have been the effective point-proton radius \cite{c12}, relative kinetic energy \cite{benchmark}, and the nucleon-nucleon (NN) pair density \cite{benchmark,npn}. An earlier application to electromagnetic operators was in a restricted model, where the effective quadrupole charges have been shown to have the values expected in a single harmonic oscillator shell, while the effective two-body contributions were found to be small \cite{navratil1997}.

The purpose of this paper is to present the formalism for general one- and two-body effective operators consistent with the effective interactions of the \textit{ab initio} NCSM. Section \ref{theory} details this approach, particularly with regard to the treatment of states outside the model space. In Sec. \ref{results} we apply the formalism to different one- and two-body operators. The renormalization is found to depend upon the range of the operator, weak for the $E2$ operator, and strong for the relative kinetic energy. Because the two-body cluster approximation used in this work integrates out mainly the short-range part of the interaction, short-range operators, such as the relative kinetic energy, are strongly renormalized, while long-range operators, e.g., quadrupole, are weakly renormalized. By using a Gaussian two-body operator of variable range, we show how much the renormalization depends on the range of the operator. We present our conclusions in Sec. \ref{concl}.

\section{Theoretical Overview}
\label{theory}

We start with a system of $A$ particles, interacting through the intrinsic Hamiltonian
\begin{equation}
H_{A}=\frac{1}{A}\sum_{i>j}\frac{(\vec p_i-\vec p_j)^2}{2m}+\sum_{i>j}V^{NN}_{ij},
\label{intrHam}
\end{equation}
where $m$ is the nucleon mass, and $V^{NN}_{ij}$ the bare NN interaction, such as the Argonne potentials in coordinate space \cite{argonne} or the non-local CD-Bonn \cite{bonn}. While realistic three-body forces have been shown to be important in obtaining the nuclear spectra \cite{marsden2002,navratil2003,GFMC3NI} and in describing electromagnetic and weak form factors \cite{hayes2003}, we consider only two-body interactions for simplicity.

The intrinsic properties of the system are not affected by the addition of the CM harmonic oscillator (HO) Hamiltonian, but by casting the new Hamiltonian in the form

\begin{eqnarray}
\lefteqn{
H_A^\Omega=H_A+\frac{\vec P^2}{2mA}+\frac{1}{2}mA\Omega^2R^2 \nonumber}\\
& &=\sum_{i=1}^A \left[ \frac{\vec{p}_i^2}{2m}
+\frac{1}{2}m\Omega^2 \vec{r}^2_i
\right] \nonumber \\
& &+ \sum_{i>j=1}^A \left[ V^{NN}_{ij}
-\frac{m\Omega^2}{2A}
(\vec{r}_i-\vec{r}_j)^2
\right] \nonumber \\
& &=\sum_{i=1}^A h_i+\sum_{i>j=1}^A v_{ij}\; ,
\label{intrCM}
\end{eqnarray}
and applying the unitary transformation on the new Hamiltonian, we improve the rate of convergence of the solution of the $A$-body problem in smaller model spaces. As we subtract the CM term in the final many-body calculation, it does not introduce any net influence on the converged intrinsic properties of the many-body calculation. Furthermore, this addition and subtraction does not affect our exact treatment of the CM motion. This procedure introduces a pseudo-dependence upon the HO frequency $\Omega$, and the two-body cluster approximation described below will sense this dependence. In the largest model spaces, however, important observables manifest a considerable independence of the frequency $\Omega$ and the model space limit.

A transformation, which accommodates the short range correlations by means of an antihermitian operator \cite{okubo,LS80,UMOA}, produces an effective Hamiltonian ${\cal H}$ given by
\begin{equation}
{\cal H}=e^{-S}H_A^\Omega e^S.
\label{transfHam}
\end{equation}
Note that even if the original Hamiltonian contained just one- and two-body terms, the operator $S$ and the transformed Hamiltonian ${\cal H}$ contain up to $A$-body terms. Obtaining the exact operator $S$ is equivalent to solving the initial problem, which would make the transformation impractical. We will discuss below, in some detail, approximations which allow us to solve the $A$-body system; for now, the derivation is exact.

The purpose of transformation (\ref{transfHam}) is to preserve the solutions of the original Hamiltonian when one reduces the  infinite dimension of the Hilbert space to a finite value which is numerically tractable. We achieve this by splitting the full space associated with the $A$-body system into $P$, or model space, and $Q$, the excluded space; the decoupling condition
\begin{equation}
Q {\cal H} P=0,
\label{decoupl}
\end{equation}
in addition to the requirements $P S P=QSQ=0$ \cite{UMOA}, ensure that correlations left out by restriction to the model space $P$ are properly taken into account for a subset of the exact eigenstates. Formally, the operator $S$ can be written \cite{UMOA} by means of another operator $\omega$ as
\begin{equation}
S=\mathrm{arctanh} (\omega-\omega^\dagger),
\end{equation}
where the new operator fulfills $Q\omega P=\omega$. The energy-independent effective Hamiltonian in the model space $P$ becomes
\begin{equation}
H_{eff}=P{\cal H}P=\frac{P +P\omega^\dagger Q}{\sqrt{P+\omega^\dagger\omega}}
H_A^{\Omega}\frac{P+Q\omega P}{\sqrt{P+\omega^\dagger\omega}},
\label{effHam}
\end{equation}
and, analogously, any arbitrary observable can be transformed to the $P$ space as \cite{okubo,navratil1993}
\begin{equation}
O_{eff}=P{\cal O}P=\frac{P +P\omega^\dagger Q}{\sqrt{P+\omega^\dagger\omega}}
O\frac{P+Q\omega P}{\sqrt{P+\omega^\dagger\omega}}\;.
\label{effOp}
\end{equation}

In order to compute effective operators one needs to know $\omega$. This operator connects eigenvectors in $P$ to vectors in the $Q$ space. A simple way to compute $\omega$ is \cite{c12}
\begin{equation}
\langle \alpha_Q|\omega|\alpha_P\rangle =\sum_{k\in {\cal K}}\langle \alpha_Q|k\rangle
\langle \tilde k|\alpha_P\rangle,
\label{omega}
\end{equation}
with $|\alpha_P\rangle$ and $|\alpha_Q\rangle$ the basis states of the $P$ and $Q$ spaces, respectively; $|k\rangle$ denotes states from a selected set ${\cal K}$ of eigenvectors of the Hamiltonian in the full space $H_A^\Omega|k\rangle=E_k|k\rangle$, and $\langle\alpha_P|\tilde k\rangle$ is the matrix element of the inverse overlap matrix $\langle\alpha_P|k\rangle$, that is $\sum_{\alpha_P}\langle k'|\alpha_P\rangle\langle \alpha_P|\tilde k\rangle=\delta_{kk'}$. Note that the dimension of the subspace ${\cal K}$ is equal with the dimension of the model space $P$.

As noted before and seen explicitly in Eq. (\ref{omega}), in order to solve for $\omega$ one needs the solution of the $A$-body problem, i.e., the eigenvectors $|k\rangle$, which is the final goal. Therefore, we introduce the cluster approximation. This consists in finding $\omega$ for the $a$-body problem, $a<A$, and then using the effective interaction thus obtained for solving the $A$-body system. This approximation introduces a real dependence of the oscillator parameter $\Omega$, and the solution to this problem is to search for a range of $\Omega$ values over which the results are weakly $\Omega$ dependent. There are two limiting cases of the cluster approximation: first, when $a\to A$, the solution becomes exact; a higher-order cluster is a better approximation and was shown to increase the rate of convergence \cite{navratil2003}. Second, when $P\to 1$, the effective interaction approaches the bare interaction; as a result, the cluster approximation effects can be minimized by increasing as much as possible the size of the model-space size. 

In this work, we present results obtained at the two-body cluster level. Under this approximation, the transformation writes as
\begin{equation}
S\approx \sum_{i>j=1}^A S_{ij},
\end{equation}
with $S_{ij}=\mathrm{arctanh}(\omega_{ij}-\omega_{ij}^\dagger)$. Applying the operator identity
\begin{equation}
e^{-S}Oe^S=O+[O,S]+\frac{1}{2!}[[O,S],S]+...
\end{equation}
to transform a general one-body operator $O^{(1)}=\sum_{i=1}^A O_i$, one obtains
\begin{equation}
{\cal O}^{(1)}=O^{(1)}+\sum_{i>j=1}^A[O_i+O_j,S_{ij}]+\sum_{i>j}^A[[O_i+O_j,S_{ij}],S_{ij}]+...,
\end{equation}
where we have retained only the one- and two-body terms, neglecting higher body contributions, such as $[O_i,S_{jk}]$, with $i\neq j$ and $i\neq k$. Resummation of the commutators yields

\begin{eqnarray}
\lefteqn{P O_{eff} P=P\sum_i O_i P\nonumber}\\
& & +P\sum_{i>j=1}^A\left[e^{-S_{ij}}\left(O_i+O_j\right)e^{S_{ij}}-
\left(O_i+O_j\right)\right]P.
\label{oneb}
\end{eqnarray}

Analogously, for a general two-body operator
\begin{equation}
P O_{eff} P=P\sum_{i>j=1}^Ae^{-S_{ij}}O_{ij}e^{S_{ij}}P,
\label{twob}
\end{equation}
and, in particular, the effective Hamiltonian derived from Eq. (\ref{intrCM}) is given by
\begin{eqnarray}
\lefteqn{P H_{eff}P=P\sum_{i=1}^Ah_i P\nonumber }\\
& &+P\sum_{i>j=1}^A \left[e^{-S_{ij}}\left( h_i+h_j+v_{ij}\right)e^{S_{ij}}-h_i-h_j\right]P.
\end{eqnarray}
We emphasize that in the two-body cluster approximation the explicit decoupling condition in Eq. (\ref{decoupl}) is fulfilled for the two-body problem now:
\[
Q_2 {\cal H}^{(2)} P_2=Q_2 e^{-S_{12}}(h_1+h_2+v_{12})e^{S_{12}} P_2=0,
\]
where $P_2$, $Q_2$ refer to the corresponding projection operators for the two-particle system. Condition (\ref{decoupl}) is, in general, violated for the $A$-body problem, but the errors become smaller with increasing the model space. The dependence upon $A$ due to the addition of the CM term in Eq. (\ref{intrCM}) is kept in $v_{12}$.

Finally, note that even if initially one starts with an one-body operator, in the two-body cluster approximation, the effective operator will generally have irreducible two-body matrix elements.

\section{Results and discussion}
\label{results}

For details regarding the procedure to obtain the effective interaction for a system of $A$ nucleons in the two-body cluster approximation, we refer the interested reader to previous work, e.g., Ref. \cite{c12}. We note that in this paper the effective interaction and the transformation $\omega$ are obtained in the relative system of two particles, in a large HO basis. The $Q$ space is chosen to be a few hundred $\hbar\Omega$ excitations in order to obtain an exact solution to the two-body Schr\" odinger equation. Due to the rotational symmetry, we formulate the problem in two-nucleon channels with good total spin $s$, total angular momentum $j$, and isospin $t$, reducing drastically the dimensions involved, when performing the summation over the states in the $Q$ space in Eq. (\ref{effHam}) The same procedure can be applied to operators which can be analytically expressed in terms of relative and CM coordinates of pairs. Therefore, we develop a convergence procedure which works for general one- and two-body operators.

As shown in the previous section, the corrections to general operators are given by Eq. (\ref{oneb}) for one-body operators, and Eq. (\ref{twob}) for two-body operators. Suppose the operators are given in the single-particle representation, so that in order to compute contributions of the form $\exp(-S_{ij})(O_i+O_j)\exp(S_{ij})$ for one-body operators [and correspondingly  $\exp(-S_{ij})O_{ij}\exp(S_{ij})$ for two-body operators] by means of Eq. (\ref{effOp}), one needs to transform either the operator to the relative system or the transformation $\omega$ to single-particle representation.  The two procedures, however, give the same result, and the numerical burden is likely comparable.

Going back to Eq. (\ref{effOp}), note that in the computation of effective operators, the number of two-body matrix elements involved in the summation over the $Q$ states becomes numerically intractable in the single-particle representation. For example, in terms of the matrix elements in the relative coordinates, the transformation $\omega$ in the single-particle picture is given by
\begin{widetext}
\onecolumngrid

\begin{eqnarray}
\lefteqn{\langle n_1l_1j_1,n_2l_2j_2; Jt |\omega|n_3l_3j_3,n_4l_4j_4;Jt\rangle
=\nonumber}\\
& & = 2\frac{1}{\sqrt{1+\delta_{n_1n_2}\delta_{l_1l_2}\delta_{j_1j_2}}}
\frac{1}{\sqrt{1+\delta_{n_3n_4}\delta_{l_3l_4}\delta_{j_3j_4}}}
\sum_{s,\Lambda,\Lambda'}\left(
\begin{array}{ccc}
l_1 & l_2 & \Lambda\\
\frac{1}{2}&\frac{1}{2} & s\\
j_1 & j_2 & J
\end{array}\right)\left(
\begin{array}{ccc}
l_3 & l_4 & \Lambda'\\
\frac{1}{2}&\frac{1}{2} & s\\
j_3 & j_4 & J
\end{array}\right)\nonumber \\
& & \times
\sum_{\stackrel{nl}{n'l'}}\sum_{NL}\langle n'l',NL;\Lambda'|n_1l_1,n_2l_2;\Lambda'\rangle
\langle nl,NL;\Lambda|n_3l_3,n_4l_4;\Lambda\rangle\nonumber\\
& & \times
\sum_jU(jLs\Lambda';Jl') U(jLs\Lambda;Jl)
\langle n'l's(j)t |\omega|nls(j)t\rangle,
\label{transf}
\end{eqnarray}

\end{widetext}
\noindent
where $|n_1l_1j_1,n_2l_2j_2;Jt\rangle$ refer to two-body states in the $Q_2$ space, and  $|n_3l_3j_3,n_4l_4j_4;Jt\rangle$ refer to two-body states in the $P_2$ space. 
We have employed the same notations as in Ref. \cite{moshinsky} for the Talmi-Moshinsky transformation.

Analyzing Eq. (\ref{transf}), we note that the quantum numbers for the CM states are restricted by the model space, $2N+L\leq N_P$, with $N_P$ fixed by the size of the model space. The relative $(n,l)$ states are restricted by the model space so that $2n+l\leq N_P$. Using the energy conservation in the Brody-Moshinsky brackets, one also obtains a restriction for the single-particles states, $2n_3+l_3+2n_4+l_4\leq N_P$. However, the states $(n',l')$ run over the excluded space and, since $2n_1+l_1+2n_2+l_2$ is not restricted by the model space, the number of possible pairs becomes numerically intractable. Hence, we restrict the relative states in the excluded space by the condition $N_P\le 2n'+l'\leq N_Q$, and observe convergence by increasing $N_Q$. (To give the reader an idea about the dimensions involved, we note that for $N_Q=28$ for a p-shell nucleus, the number of $\omega$ matrix elements, taking into account the possible symmetries, in a $2\hbar\Omega$ model space defined by $N_P=4$, is 413,163; subsequently, the number of reduced matrix elements involved in the transformation of a tensor operator with $J=1$, $T=1$ is 7,216,180.) This procedure was successfully tested for the deuteron in a restricted space, where we have shown that, for general tensor operators, the matrix elements obtained with effective operators approach the values obtained with bare operators in the full space \cite{brucefest}. 

\begin{figure}
\includegraphics*[scale=0.48]{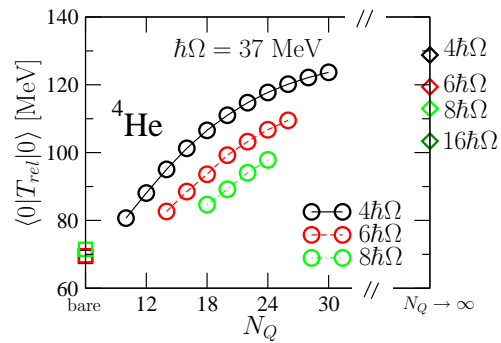}
\caption{(Color online) The expectation value of the relative kinetic energy on the ground state using realistic wave functions for $^{4}$He in different model spaces. We show how the ground state expectation value changes when one increases the number of states in the $Q$-space (circles). For comparison, we include the results obtained with bare operator (squares) and effective operator including all states in the excluded space (diamonds).}
\label{KE}
\end{figure}

In order to test the convergence procedure in a realistic model space, one could consider two-body operators which depend only upon the relative coordinates, which can be renormalized similarly to the Hamiltonian. One such operator is the relative kinetic energy, whose effective matrix elements can be computed by simply replacing in the effective interaction code the bare Hamiltonian with the bare kinetic energy after the transformation $\omega$ has been determined. The results summarized in Fig. \ref{KE} show that the expectation value computed with the approximate effective operator converges toward the effective value computed including all the states in the excluded space. While the convergence rate might look slow, we would like to point out that even for small values of $N_Q$ we obtain reasonable renormalization comparing to the full space renormalization, in contrast to the interaction, where several hundred $\hbar\Omega$ excitations are usually necessary. Note that larger model spaces require less renormalization with respect to the bare operator, as expected. Also, the effective expectation values are similar to previous results \cite{benchmark}, although the latter have been obtained with wave functions computed in the three-body cluster approximation in $16\hbar\Omega$. This can be explained by the character of the kinetic energy operator which is zero range, so that it is well renormalized at the two-body cluster level. 

\begin{figure}
\includegraphics*[scale=0.6]{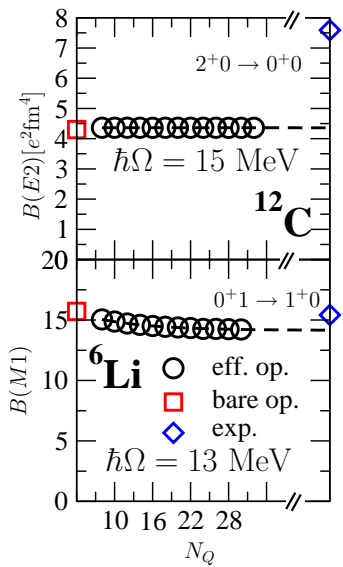}
\caption{(Color online) $2\hbar\Omega$ model space results for $B(E2)$ in $^{12}$C and $B(M1)$ in $^{6}$Li using effective interaction derived from AV8' potential. We show the results as a function of the dimension of the $Q$-space included. For comparison, we included the values obtained using the bare operators (squares) and the experimental values (diamonds). The dashed lines represent an interpolation of the strengths obtained with effective operators.}
\label{C12bval}
\end{figure}

Our first applications of the formalism to one-body operators are to quadrupole and $M1$ transition operators. We present in Fig. \ref{C12bval} the $2\hbar\Omega$ model space results for $B(E2; 2^+_1 0\to 0^+_1 0)$ in $^{12}$C and $B(M1; 1^+_1 0\to 0^+_1 0)$ in $^6$Li. The shell model calculations have been performed using the many-fermion dynamics code \cite{MFD}. Following the procedure described above, we compute the effective operator by means of Eq. (\ref{oneb}) by adding from the $Q$ space two-body matrix elements  $(2n'+l')\leq N_Q$, and increasing $N_Q$. Note that for $E2$ transitions we expect the biggest contribution to come from $4\hbar\Omega$, as the $E2$ operator connects across two shells. Figure \ref{C12bval} shows, however, that $B(E2; 2^+_1 0\to 0^+_1 0)$ remains essentially flat, at the same value as the one obtained with the bare operator, and at about half the experimental strength. We have included the experimental $B(E2)$ for reference.  What we do not have is the exact results for the full Hilbert space which would provide the ultimate comparison with experiment.  Instead, we present the convergence with increasing $N_Q$ at the 2-body cluster level as shown in Fig. \ref{C12bval}.  Our achievement here is to include the contribution for renormalization at the two-body cluster level which turns out to be rather small. The large residual discrepancy between our extrapolated result and experiment is due to the combination of residual effective three-body effects and to the neglect of genuine three nucleon potentials. However, we argue below that the result obtained with only two-body interactions should be significantly closer to the experiment.

For the $M1$, we observe a small correction in the transition strength, albeit the correction is slightly larger than the discrepancy between theory and experiment. For other $M1$ transitions, however, we observe, in general, the same minimal effect as for quadrupole transitions. Such a small correction can be more easily understood, as $M1$ does not connect the model space with the complementary one; increasing the model space introduces enough correlations in the wave function so that using the bare operator gives $B(M1)$ values close to experiment.

Despite the successful test in the case of the relative kinetic energy, one can still ask the question whether the convergence procedure is faulty and if including all the states in the excluded space in calculations would not improve the $E2$ transition strengths. Because, in general, the electric multipoles can be written as sum of terms which factorize the CM and relative contributions for each pair of nucleons, we can include all the states in the $Q$-space. (The corresponding transition strengths are analogous to the $N_Q\to \infty$ values in Fig. \ref{KE}.) In the Appendix, we identify the CM and relative contributions for each pair to the $E2$ operator. In this context, we further show that the difference between the usual $E2$ operator used in shell model calculations, and the translationally invariant one is a term which involves only contributions from the CM of the $A$-body system. The latter is irrelevant when employing translationally invariant wave functions or many-body wave functions which factorizes exactly in intrinsic and a $0\hbar\Omega$ CM contributions, as in our case.

In Table \ref{BE2Li6}, we present the $B(E2)$ values for quadrupole transitions in a $2\hbar \Omega$ model space in $^6$Li, and compare them with the experimental strengths, where available. In these calculations we include in the summation all the states in the $Q$ space. Because, as noted before, we limit ourselves to small model spaces, the transition strength obtained with bare operators are far from the experimental values. As noted before, part of the discrepancies can appear because of the missing three-body forces in our calculations. However, we argue that the theoretical strength obtained in the model which includes only two-body forces should be closer to experiment. Thus, increasing the model space up to $10\hbar\Omega$ reduces significantly the discrepancy between theory and experiment for the $1^+0\rightarrow 3^+0$ transition in Table \ref{BE2Li6}. A similar trend is observed in larger model spaces accessible today, which suggests that indeed the strength in the full space should be significantly closer to experiment even in the absence of three-body interactions. Note, however, the large difference between the strengths obtained with bare operators in $2\hbar\Omega$ and $10\hbar\Omega$. Because of the convergence properties discussed for the effective interaction in the previous section, which can be extended to other operators, one expects that the bare value in the larger model space would be closer to the exact theoretical value than the one in the smaller space. Therefore, one may conclude that at least the value in the $2\hbar\Omega$ space is far from the correct value, and that the effective operator should have a significant impact in this model space.  Table \ref{BE2Li6} shows the opposite result, \textit{i.e.}, very little difference between the transition strengths obtained with bare and effective operators. In contrast, for the deuteron, where the two-body cluster provides the exact solution, the bare quadrupole operator in $4\hbar\Omega$ gives 0.179 $e$ fm$^2$ for the quadrupole moment, while the value of 0.270 $e$ fm$^2$, described by the AV8' potential, is obtained using the corresponding effective operator. Overall, the difference between the bare operator results in the $2\hbar\Omega$ and $10\hbar\Omega$ model spaces, coupled with the small renormalization at the two-body cluster level, indicate there are sizable effective multi-body interaction effects needed to correct the $2\hbar\Omega$  $B(E2)$ value.

\begin{table}
\caption{$B(E2)$ values, in $e^2$ fm$^4$, for $^{6}$Li computed with bare and effective operators in a $2\hbar\Omega$ model space ($\hbar\Omega=13$ MeV). All states in the $Q$ space are included (see text for details). For comparison we have included also $B(E2)$ values obtained in $10\hbar\Omega$ space with the bare quadrupole operator \cite{spectr6}. Experimental values are from Ref. \cite{e2exp}.}
\label{BE2Li6}
\begin{ruledtabular}
\begin{tabular}{ccccc}
 & \multicolumn{2}{c}{$2\hbar \Omega$} & $10\hbar \Omega$ & Expt. \\
 \cline{2-3}
 & Bare & Effective & Bare  & \\
\hline
 $1^+0\rightarrow 3^+0$ & 2.647 & 2.784 & 10.221 & 21.8(4.8)\\
 $2^+0\rightarrow 1^+0$ & 2.183 & 2.269 & 4.502 &4.41(2.27)\\
 $1_2^+0\rightarrow 1^+0$ & 3.183 & 3.218 & &
\end{tabular}
\end{ruledtabular}
\end{table}

Because the two-body cluster approximation accommodates the short range correlations, one can expect that such an approach might not be well-suited to renormalize the $E2$ operator, which is infinite range. To illustrate the importance of the range of the operator for renormalization at the two-body cluster level, we consider a Gaussian two-body operator of range $a_0$

\begin{equation}
O(\vec r_1,\vec r_2)=C_0\exp\left(-\frac{(\vec{r}_1-\vec{r}_2)^2}{a_0^2}\right),
\label{gaussopdef}
\end{equation}
with $C_0$ chosen so that
\[
C_0\int d\vec{r}\:\: \exp\left(-\frac{r^2}{a_0^2}\right)=1.
\]
While this is not a realistic observable, one can often expand realistic operators as sums of Gaussians, so this could be used to estimate the renormalization of different contributions. However, in this paper, the only purpose of this example is to illustrate the dependence of the renormalization upon the range of the operator. We define the renormalization as $(\langle O_{eff} \rangle-\langle O_{bare}\rangle)/ \langle O_{bare}\rangle$, and in Fig. \ref{gaussop}(a) we summarize the results using the realistic ground-state wave function for $^4$He. At small ranges, the expectation value computed with the effective operator is significantly different from the one obtained using the bare operator. However, when the operator becomes longer range, the renormalized value becomes nearly indistinguishable from the bare value. But in the absence of an exact full space expectation value of this operator one cannot tell whether or not the absence of renormalization for long range operators is a consequence of the fact that the bare operator is already close to the exact result. However, in Fig. \ref{gaussop}(b) we show the dependence of the ground-state expectation value of the operator upon the size of the model space for selected short- and long-range operators. For the short-range operator ($a_0=0.2$ fm), the expectation value obtained with the bare operator varies with the space, while the one obtained with the effective operator is flat, suggesting that indeed the calculation is converged in this case. On the other hand, the expectation value of the long-range operator ($a_0=1$ fm) presents about the same dependence on the model-space size with both bare and effective operators, therefore suggesting that the effective operator at this range is as poor an approximation to the exact result of the full space as the bare operator - a feature reminiscent of our $B(E2)$ results above. That is, while we do not expect realistic operators to behave exactly as the Gaussian operator used here, we believe that this example offers a qualitative understanding of the very weak renormalization of the quadrupole operator.

\begin{widetext}
\onecolumngrid

\begin{figure}
\includegraphics*[scale=0.7]{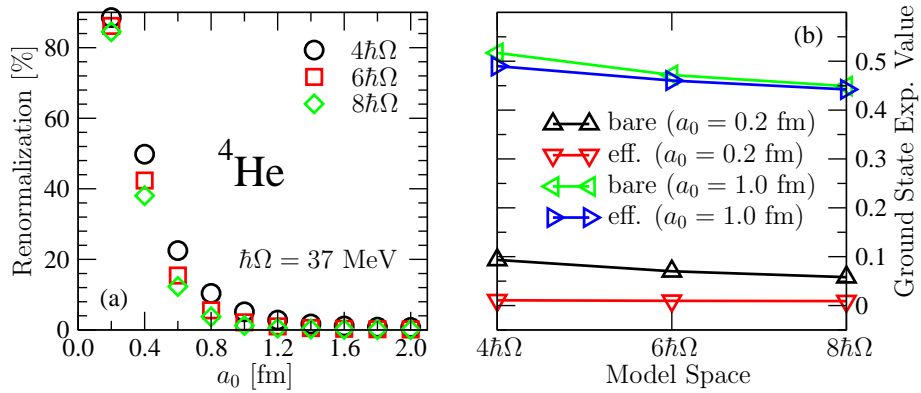}
\caption{(Color online) Left panel: renormalization of the ground state expectation value of the relative Gaussian operator using realistic wave functions for $^{4}$He, as a function of the range of the operator for $4\hbar\Omega$ (circles), $6\hbar\Omega$ (squares), and $8\hbar\Omega$ (diamonds). Right panel: expectation values in different model spaces for Gaussian operators of selected ranges. All the states in the $Q$-space are included.}
\label{gaussop}
\end{figure}
\end{widetext}

\section{Summary and Conclusions}
\label{concl}

We have computed the effective operators consistent with the \textit{ab initio} NCSM effective interactions. In the present investigation we limit ourselves to solutions obtained in the two-body cluster approximation. The main purpose of this paper was a qualitative investigation of the effective operators rather than a highly accurate description of the experimental data; this is the reason we have considered limited shell model spaces for $^4$He, $^6$Li and $^{12}$C nuclei. Such small model spaces offer an excellent testing ground for effective operators, for one expects that the smaller the model space, the more significant the renormalization.

For general operators one cannot apply the same calculational procedure as for the Hamiltonian; for these we have developed a convergence procedure in which we add correlations from the excluded space a few shells at a time. This procedure can be applied to arbitrary one- and two-body operators. We have tested this procedure in the case of the relative kinetic energy, where we obtained slow convergence as the number  of $Q$-states included in the renormalization is increased.

We found that the quadrupole operator is very weakly renormalized at the two-body cluster level. We suggest that this is a consequence of the character of multipole operators which have infinite range. To substantiate this hypothesis, we have shown, using a Gaussian operator that the renormalized operator is quite sensitive to the range of the operator. Thus, the shorter the range, the stronger the renormalization, with only a small effect for long range operators. Therefore, in order to describe long range operators one needs to go beyond the two-body cluster approximation. As a general caveat, any truncation of the space could induce effective operators with non-negligible higher-body correlations. This result is in accord with previous findings of tests in restricted models for the double-$\beta$ decay operator \cite{engel2004}. It should be noted that earlier calculations for $^6$Li, which obtained significant effective quadrupole charge renormalization, were based on 
large-basis NCSM calculations, which were then explicitly truncated into
a $0\hbar\Omega$ space and fitted to one-plus-two-body quadrupole operators \cite{navratil1997}.
By construction these calculations contained all correlations up to
six-body due to the truncation and, hence, yielded the large effective
charge renormalizations that are found experimentally. Techniques for including these high-body correlations in our calculations are under investigation.

Nevertheless, even at the two-body cluster level, there are cases where the renormalization is significant, as we have shown for the relative kinetic energy. Future work will investigate electromagnetic processes, where one expects the two-body cluster renormalization to play an important role at large momentum transfer.

\acknowledgments

I.S. and B.R.B acknowledge partial support by NFS grants PHY0070858 and PHY0244389.
The work was performed in part under the auspices of the U. S. Department of Energy by the University of California, Lawrence Livermore National Laboratory under contract No. W-7405-Eng-48. P.N. received support from LDRD contract 04-ERD-058. J.P.V. acknowledges partial support by USDOE grant No DE-FG-02-87ER-40371.  We thank the Institute for Nuclear Theory at the University of Washington for its hospitality and the Department of Energy for partial suport during during the completion of this work.\\

\appendix*
\section{The translationally invariant quadrupole operator}

In this Appendix we revisit the quadrupole operator, identifying the intrinsic and CM contributions.

We start by rewriting the single-particle isoscalar quadrupole operator in a two-body form, that is
\begin{equation}
E2=\sum_i  O(\vec r_i)=\frac{1}{A-1}\sum_{i>j=1}^A\left(O(\vec r_i)+O(\vec r_j)\right),
\label{e2def}
\end{equation}
where $O(\vec r_i)=e_{IS} r_i^2 Y_2(\hat r_i)$, with $Y_2$ the spherical harmonics of rank 2 and $e_{IS}$ the isoscalar charge. For a pair of particles ($i$, $j$) we define the relative $\vec r_{ij}=\vec r_i-\vec r_j$ and the CM $\vec R_{ij}=(\vec r_i+\vec r_j)/2$ coordinates. Using Eq. (35) in Ref. \cite{varsh}, one can show that
\begin{equation}
O(\vec r_i)+O(\vec r_j)=\frac{1}{2}O(\vec r_{ij}) + 2 O(\vec R_{ij}).
\label{E2icm}
\end{equation}
This separates for a pair of particles the relative and CM contributions to the quadrupole operator. The Lee-Suzuki transformation affects only the relative coordinates, so that in this form one can apply a similar procedure as for the renormalization of the Hamiltonian. Note, however, that because the quadrupole operator is a tensor operator of rank 2, it can couple different channels with different total angular momentum $j$, unlike the Hamiltonian. Hence, the renormalization procedure for the $E2$ multipole is more involved than for the Hamiltonian.

We can show in general that any electric multipole operator can be written in a form that allows similar renormalization as for the Hamiltonian. Thus, consider the identities:
\begin{subequations}
\begin{equation}
\exp(i\,\vec q\cdot \vec r_i)=\exp(i\,\vec q\cdot \vec r_{ij}/2)\exp(i\,\vec q\cdot \vec R_{ij})
\end{equation}
\begin{equation}
\exp(i\,\vec q\cdot \vec r_j)=\exp(-i\,\vec q\cdot \vec r_{ij}/2)\exp(i\,\vec q\cdot \vec R_{ij}).
\end{equation}
\end{subequations}
Using the usual expansion of the exponentials in spherical Bessel functions and spherical harmonics, one can show that
\begin{widetext}
\onecolumngrid
\begin{eqnarray}
\lefteqn{
i^L (j_L(qr_i)Y_L (\hat r_i)+j_L(qr_j)Y_L (\hat r_j))=\nonumber } \\
& & \sqrt{4\pi}
\sum_{ll'}\sqrt{\frac{(2l+1)(2l'+1)}{2L+1}}i^{l+l'}(1+(-1)^l)\langle l 0, l' 0|L 0\rangle j_l(qr_{ij}/2)j_{l'}(qR_{ij})
[Y_l(\hat r_{ij})\otimes Y_{l'}(\hat R_{ij})]_L.
\end{eqnarray}
In the limit of zero momentum transfer, we indeed find that the electric multipoles can be written as sum of terms which factorize the relative and CM contributions of nucleon pairs. In particular, for $L=2$, we regain Eq. (\ref{E2icm}).

An important issue in shell model calculations is separation of intrinsic and CM excitations. Although our wave functions factorize exactly in intrinsic and CM contributions, one might pick up spurious contributions if one does not use translationally invariant operators. In the following, we concentrate on the quadrupole operator. Because the last term in Eq. (\ref{E2icm}) cannot be summed in a part involving only the $A$-body CM, one might expect that the CM can introduce spurious contributions. We point out, however, that the CM of individual pairs can be changed without changing the CM of the $A$-body system. Furthermore, we can show that there is no contribution from the $A$-body CM. Thus, we start with the translationally invariant expression of the quadrupole operator
\begin{equation}
E2=\sum_i O(\vec r_i -\vec R_{CM}),
\label{e2ti}
\end{equation}
so that Eq. (\ref{E2icm}) becomes
\begin{equation}
O(\vec r_i-\vec R_{CM})+O(\vec r_j-\vec R_{CM})=\frac{1}{2}O(\vec r_{ij}) + 2 O(\vec R_{ij}-\vec R_{CM}).
\end{equation}
This can be further transformed by means of Eq. (35) in Ref. \cite{varsh}:
\begin{eqnarray}
\lefteqn{
O(\vec r_i-\vec R_{CM})+O(\vec r_j-\vec R_{CM})= \nonumber}\\
& & \frac{1}{2}O(\vec r_{ij}) +
2O(\vec R_{ij})+2O(\vec R_{CM})-\frac{\sqrt{ 4 \pi 5!}}{3}[R_{ij} Y_1 (\hat R_{ij}) \otimes R_{CM} Y_1(\hat R_{CM})]_2.
\label{e2cm}
\end{eqnarray}

\noindent
Because $\vec R_{ij}=R_{ij}Y_1(\hat R_{ij})$ one can sum contributions from all pairs $(i,j)$, so that Eq. (\ref{e2cm}) becomes:
\begin{equation}
\sum_i O(\vec r_i-\vec R_{CM}) =\sum_i O(\vec r_i)+R_{CM}^2 Y_2(\hat R_{CM})-\frac{\sqrt{4 \pi 5!}}{3}A(A-1)R_{CM}^2[Y_1(\hat R_{CM})\otimes Y_1(\hat R_{CM})]_2.
\end{equation}

\end{widetext}

Therefore, when computing $E2$ transition strengths with wave functions which factorize the intrinsic and $0\hbar\Omega$ CM pieces, there are no spurious contributions even though the operator usually employed, Eq. (\ref{e2def}), is not translationally invariant. The latter differs from its translationally invariant form by a term which contains only an irrelevant tensor contribution from the CM of the $A$-body system.


\begin{thebibliography}{99}
\bibitem{okubo}  S. Okubo, Prog. Theor. Phys. \textbf{12}, 603 (1954).
\bibitem {LS80} J. Da Providencia and C. M. Shakin, Ann. of Phys. \textbf{30}, 95 (1964); K. Suzuki and S.Y. Lee, Prog. Theor. Phys. {\bf 64}, 2091 (1980); K. Suzuki, Prog. Theor. Phys. {\bf 68}, 246 (1982); K. Suzuki and R. Okamoto, Prog. Theor. Phys. {\bf 70}, 439 (1983).
\bibitem{UMOA}  K. Suzuki, Prog. Theor. Phys. {\bf 68}, 1999 (1982);
K. Suzuki and R. Okamoto, Prog. Theor. Phys. {\bf 92}, 1045 (1994).
\bibitem{spectr6} P. Navr\'atil, J.~P. Vary, W.~E. Ormand, and B.~R. Barrett, Phys. Rev. Lett. \textbf{87}, 172502 (2001).
\bibitem{c12lett} P. Navr\' atil, J. P. Vary, and B. R. Barrett, Phys. Rev. Lett. \textbf{84}, 5728 (2000).
\bibitem{c12} P. Navr\' atil, J. P. Vary, and B. R. Barrett, Phys. Rev. C \textbf{62},
054311 (2000).
\bibitem{benchmark} H. Kamada, et. al., Phys. Rev. C \textbf{64}, 044001 (2001).
\bibitem{npn} B. R. Barrett, B.~Mihaila, S.~C.~Pieper, and R.~B.~Wiringa, Nucl. Phys. News \textbf{13}, No. 1, 17 (2003).
\bibitem{navratil1997} P. Navr\' atil, M. Thoresen, and B. R. Barrett, Phys. Rev. C \textbf{55}, R573 (1997).
\bibitem {argonne} R.~B. Wiringa, V.~G.~J. Stoks and R. Schiavilla,
                Phys. Rev. C {\bf 51}, 38 (1995); 
               B. S. Pudliner, V. R. Pandharipande, J. Carlson,
               S. C. Pieper and R. B. Wiringa,
               Phys. Rev. C {\bf 56} 1720, (1997);
               R. B. Wiringa, Nucl. Phys. {\bf A 631}, 70c (1998);
               S. Pieper and R. B. Wiringa, 
               Annu. Rev. Nucl. Part. Sci. 51, 53 (2001).
\bibitem{bonn} R. Machleidt, F. Sammarruca and Y. Song, Phys. Rev. C
                 {\bf 53}, R1483 (1996);
                 R. Machleidt, Phys. Rev. C {\bf 63}, 024001 (2001).
\bibitem{navratil2003} P. Navr\' atil and E. W. Ormand, Phys. Rev. C \textbf{68}, 034305 (2003).
\bibitem{marsden2002} D.~C.~J.~Marsden, P. Navr\' atil, S.~A.~Coon and B.~R.~Barrett, Phys. Rev. C \textbf{66}, 044007 (2002).
\bibitem{GFMC3NI}S. C. Pieper, K. Varga, and R.~B. Wiringa, Phys. Rev. C \textbf{66}, 
044310 (2002).
\bibitem{hayes2003} A. C. Hayes, P. Navr\'atil, and J. P. Vary, Phys. Rev. Lett. \textbf{91}, 012502 (2003).
\bibitem{navratil1993}P. Navr\'atil, H. Geyer, and T.~T.~S.~Kuo, Phys. Lett. \textbf{B 315}, 1 (1993).
\bibitem{moshinsky}M. Moshinsky, \textit{The harmonic oscillator in modern physics; from atoms to quarks} (Gordon and Breach, New York, 1969).
\bibitem{MFD} J. P. Vary, \textit{The Many-Fermion Dynamics Code}, Iowa State University (1992); J. P. Vary and D. C. Zheng, \textit{ibid.} (1994) (unpublished).
\bibitem{brucefest} I. Stetcu, B. R. Barrett, P. Navr\'atil, and C. W. Johnson, submitted to Int. J. Mod. Phys. E [arXiv:nucl-th/0409072].
\bibitem{e2exp} F. Ajzenberg-Selove, Nucl. Phys. \textbf{A490}, 1 (1988).
\bibitem{engel2004} J. Engel and P. Vogel, Phys. Rev. C \textbf{69}, 034304 (2004).
\bibitem{varsh}D. A. Varshalovich, A.~N.~Moskalev, and V.~K.~Khersonskii, \textit{Quantum Theory of Angular Momentum} (World Scientific, Singapore, 1988), pp. 167.

\end{thebibliography}
\end{document}